\documentclass[11pt, oneside]{article}   	
\usepackage{geometry}                		
\usepackage{graphicx}				
\usepackage{amssymb}
\usepackage{amsmath}
\usepackage{subfig}
\usepackage{caption}
\usepackage{comment}
\usepackage{amsthm,verbatim,amsfonts,amscd,float,physics}
\usepackage{appendix}

\usepackage[utf8]{inputenc}
\usepackage{xcolor}
\numberwithin{equation}{section}
\begin{document}
\title{$Z_3$ gauge theory coupled to fermions and quantum computing}
\author{Ronak Desai$^{(1)}$, Yuan Feng$^{(2)}$, Mohammad Hassan$^{(3)}$, \\
Abhishek Kodumagulla$^{(4)}$, Michael McGuigan$^{(5)}$\\
(1) Rowan University current address Ohio State University, (2) Pasadena City College, \\
(3) City College of New York, 
(4) Columbia University, (5) Brookhaven National Laboratory}
\date{}
\maketitle
\begin{abstract}
We study the $Z_3$ gauge theory with fermions  on the quantum computer using the Variational Quantum Eigensolver (VQE) algorithm with  IBM QISKit software. Using up to 9 qubits we are able to obtain accurate results for the ground state energy. Introducing nonzero chemical potential we are able to determine the Equation of State (EOS) for finite density on the quantum computer. We discuss possible realizations of quantum advantage for this system over classical computers with regards to finite density simulations and the fermion sign problem.
\end{abstract}
\newpage

\section{Introduction}

Discrete gauge theories are excellent models to study on a quantum computer as they can be realized with fewer qubits than gauge theories with continuous groups like $SU(2)$ or $SU(3)$ \cite{Banuls:2016gid}
\cite{Pisarski:2021aoz}
\cite{Frank:2019jzv}
\cite{Borla:2019chl}
\cite{Borla:2020upy}\cite{Shapere:1988zv}
\cite{Gorantla:2020jpy}
\cite{Gorantla:2021svj}. When coupled to fermions at finite density and nonzero chemical potential they can have a sign problem which presents an opportunity for an advantage for quantum computing over classical computing. Also one can study time evolution for the discrete gauge models which is also difficult to simulate classically. In this paper we study the $Z_3$ gauge model coupled to fermions. This model will allow us to study the aspect of discrete gauge models described above that can potentially lead to quantum advantage.

This paper is organized as follows. In section two we describe the Hamiltonian for the $Z_3$ gauge model and how to represent the variables in terms of qubits. We also describe the Variational Quantum Eigensolver (VQE)  algorithm and how it can be applied to the $Z_3$ gauge model.  We describe the simplest configuration of two fermion vertices and one gauge link. In section three we describe the calculation for three fermion vertices and two gauge links and in section four we discuss a system with three fermion vertices and three gauge links. In section five we introduce a chemical potential and describe the equation of state at finite density. In section seven we state the main conclusions for the paper.

\section{VQE for One boson link and two fermion vertices } 

The discrete $Z_3$ gauge model coupled to fermions with gauge links defined by$A_J$, the canonically conjugate electric field variables $E_j$ and fermions $c_j$ is decribed by the Hamiltonian \cite{Banuls:2016gid}: 
\begin{equation}
H = \sum\limits_j {\frac{1}{2}} E_j^2 +  m\sum\limits_j {{{( - 1)}^j}c_j^\dag {c_j}}  + \mu \sum\limits_j {c_j^\dag {c_j}}  + \frac{i}{2}( \sum\limits_j {{e^{i g \frac{2 \pi}{3}{A_j}}}c_{j + 1}^\dag {c_j}}- h.c.) 
\end{equation}
The above Hamiltonian is realized in terms of staggered fermions as  in studies of the Schwinger model \cite{Banks:1975gq}
\cite{Shaw:2020udc}
\cite{Klco:2018kyo} although on can also consider Wilson fermions as in \cite{Zache:2018jbt}. For the simplest case of two fermion sites and one link we define the discretized operators in terms of qubits for the fermions as:
\[{c_1} = \left( {\begin{array}{*{20}{c}}
0&1\\
0&0
\end{array}} \right) \otimes \left( {\begin{array}{*{20}{c}}
1&0\\
0&1
\end{array}} \right) \otimes {I_3}\]
\begin{equation}
{c_2} = \left( {\begin{array}{*{20}{c}}
1&0\\
0&{ - 1}
\end{array}} \right) \otimes \left( {\begin{array}{*{20}{c}}
0&1\\
0&0
\end{array}} \right) \otimes {I_3}
\end{equation}
\newpage
\noindent and for the boson link as:
\[{A_1} = {I_4} \otimes X\]
\begin{equation}{E_1} = {I_4} \otimes P\end{equation}
where :
\begin{equation}X = \left( {\begin{array}{*{20}{c}}
{ - 1}&0&0\\
0&0&0\\
0&0&1
\end{array}} \right)\end{equation}
and
\begin{equation}P = {S^\dag }XS\end{equation}
with $S$ the Sylvester matrix given by:
\begin{equation}S = \frac{1}{{\sqrt 3 }}\left( {\begin{array}{*{20}{c}}
{{e^{2\pi i/3}}}&1&{{e^{ - 2\pi i/3}}}\\
1&1&1\\
{{e^{ - 2\pi i/3}}}&1&{{e^{2\pi i/3}}}
\end{array}} \right)\end{equation}
Defined this way the Hamiltonian is a $12\times 12$ matrix but one can easily adjoin a $4\times 4$ identity to it to make the Hamiltonian $16 \times 16$ and represent it in terms of 4 qubits.

To determine the ground state of the Hamiltonian on a quantum computer we use the Variational Quantum Eigensolver alogorithm (VQE). The VQE is a hybrid quantum-classical algorithm based on the variational method of quantum mechanics. One ontains an upper bound on the ground state energy by minimizing the expression:
\begin{equation}{E_0}({\theta _i}) = \frac{{\left\langle {\psi ({\theta _i})} \right|H\left| {\psi ({\theta _i})} \right\rangle }}{{\left\langle {\psi ({\theta _i})} \right|\left. {\psi ({\theta _i})} \right\rangle }}\end{equation}
The parameters of the variational wave function $\theta_i$ are angles associated with Unitary gates the can be realized on a quantum computer. The Hamiltonian can be expanded in terms a sum of tensor product of the the three Pauli matrices plus the Identity matrix which can also be represented on the quantum computer.The minimization over the angles $\theta_i$ is done using an optimizer running on a classical computer. This is what the VQE algorithm is a classical-quantum algorithm as dome parts of the computation are done on the quantum computer and others are done using classical computing. More details on the implementation of the Variational Quantum Eigensolver algorithm can be found in \cite{Kandala}. 

For the 4 qubit Hamiltonian we found using the VQE and the Sequential Least SQuares Programming (SLSQP) an upperbound on the ground state energy with results in table 1   which are close to the exact value determined by exact diagonalization. We used the state-vector simulator in IBM QISKit for the simulations in this paper. Other simulators available in QISKit include the QASM simulator with the ability to simulate noise on a quantum computer. Other backends beside the simulator include various hardware quantum computer implementations using superconducting qubits available from IBM. 
In addition to the ground state energy the VQE also determines an final variational form which can be taken as an approximation to the ground state wave function. On a quantum computer this is represented as a quantum circuit.
    \begin{figure}
    \centering
    \begin{minipage}[b]{0.5\linewidth}
      \centering
      \includegraphics[width=\linewidth]{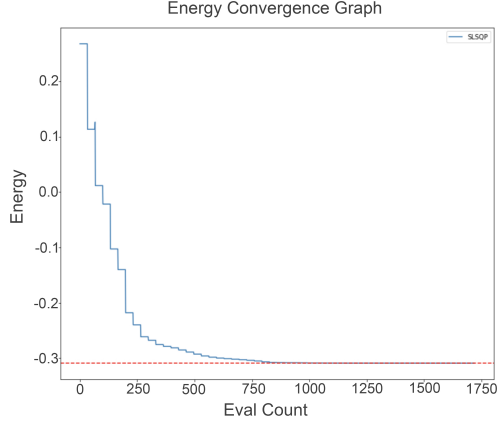}
    \end{minipage}
    \caption{Convergence graph for one boson link and two fermion vertices.}
    \label{3-qubit_unit}
\end{figure}
\begin{table}[ht]
\centering
\begin{tabular}{|l|l|l|l|l|}
\hline
Hamiltonian       & Qubits & Paulis  & Exact Result & VQE Result \\ \hline
One  $Z_3$ boson Two fermions  & 4 & 20 & -0.30901699 &  -0.30900573\\ 
 \hline
\end{tabular}
\caption{\label{tab:BasisCompare}  VQE results for $Z_3$ gauge model  model with one boson links and two fermion vertices with gauge coupling .15.  The Hamiltonian was mapped to 4-qubit operators with 20 Pauli terms.  The quantum circuit for each simulation utilized an \(R_y R_z\) variational form, with a fully entangled circuit of depth 3. The backend used was a statevector simulator. The Sequential Least SQuares Programming (SLSQP) optimizer was used, with a maximum of 600 iterations.}
\end{table}
\section{VQE for Two boson links and three fermion vertices}

For two boson links and three fermion vertices we can proceed similarly. We define three fermion annihilation operators as:
\[{c_1} = \left( {\begin{array}{*{20}{c}}
0&1\\
0&0
\end{array}} \right) \otimes \left( {\begin{array}{*{20}{c}}
1&0\\
0&1
\end{array}} \right) \otimes \left( {\begin{array}{*{20}{c}}
1&0\\
0&1
\end{array}} \right) \otimes {I_9}\]
\[{c_2} = \left( {\begin{array}{*{20}{c}}
1&0\\
0&{ - 1}
\end{array}} \right) \otimes \left( {\begin{array}{*{20}{c}}
0&1\\
0&0
\end{array}} \right) \otimes \left( {\begin{array}{*{20}{c}}
1&0\\
0&1
\end{array}} \right) \otimes {I_9}\]
\begin{equation}{c_3} = \left( {\begin{array}{*{20}{c}}
1&0\\
0&{ - 1}
\end{array}} \right) \otimes \left( {\begin{array}{*{20}{c}}
1&0\\
0&{ - 1}
\end{array}} \right) \otimes \left( {\begin{array}{*{20}{c}}
0&1\\
0&0
\end{array}} \right) \otimes {I_9}\end{equation}
\newpage
The operators associated with the two bosonic links are:
\[{A_1} = {I_8} \otimes X \otimes {I_3}\]
\[{A_2} = {I_8} \otimes {I_3} \otimes X\]
\[{E_1} = {I_8} \otimes P \otimes {I_3}\]
\begin{equation}{E_2} = {I_8} \otimes {I_3} \otimes P\end{equation}
The Hamiltonian is represented by a $72 \times 72$ matrix. One can adjoin onto it a $56\time56$ Identity matrix to make the Hamiltonian $128 \times 128$ and represent it on the quantum computer in terms of seven qubits. Running the VQE algorithm with the SLQP optimizer we find  the results in table 2  which are in close agreement with the value found from exact diagonalization.
    \begin{figure}
    \centering
    \begin{minipage}[b]{0.5\linewidth}
      \centering
      \includegraphics[width=\linewidth]{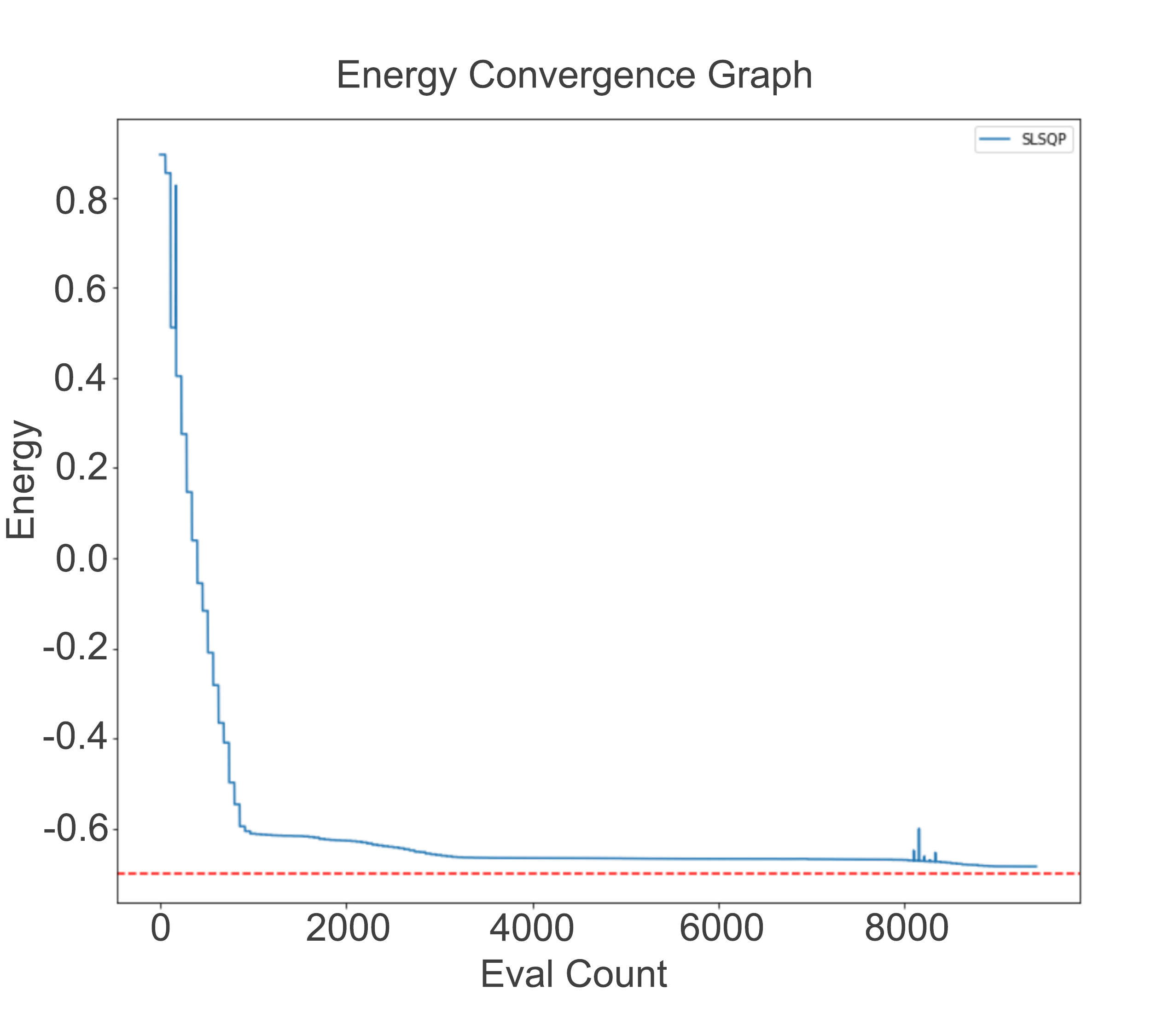}
    \end{minipage}
    \caption{Convergence graph for two boson links and three fermion vertices.}
    \label{3-qubit_unit}
\end{figure}
\begin{table}[ht]
\centering
\begin{tabular}{|l|l|l|l|l|}
\hline
Hamiltonian       & Qubits & Paulis  & Exact Result & VQE Result \\ \hline
Two $Z_3$ bosons Three fermions  & 7 & 196 & -0.69930137 & -0.68434208\\ 
 \hline
\end{tabular}
\caption{\label{tab:BasisCompare}  VQE results for $Z_3$ gauge model  model with two boson links and three fermion vertices with gauge coupling .15.  The Hamiltonian was mapped to 7-qubit operators with 196 Pauli terms.  The quantum circuit for each simulation utilized an \(R_y R_z\) variational form, with a fully entangled circuit of depth 3. The backend used was a statevector simulator. The Sequential Least SQuares Programming (SLSQP) optimizer was used, with a maximum of 600 iterations.}
\end{table}

\section{VQE for three boson links and three fermion vertices}

For the three fermion vertices and three gauge boson links we have a closed structure of a triangle.  We can use the same fermion operator representation as the previous section and define operators for the three bosons given by:
\[{A_1} = {I_8} \otimes X \otimes {I_3} \otimes {I_3}\]
\[{A_2} = {I_8} \otimes {I_3} \otimes X \otimes {I_3}\]
\[{A_3} = {I_8} \otimes {I_3} \otimes {I_3} \otimes X\]
\[{E_1} = {I_8} \otimes P \otimes {I_3} \otimes {I_3}\]
\[{E_2} = {I_8} \otimes {I_3} \otimes P \otimes {I_3}\]
\begin{equation}{E_3} = {I_8} \otimes {I_3} \otimes {I_3} \otimes P\end{equation}
The Hamiltonian is represented by a $216 \times 216$ matrix. One can adjoin onto it a $40\time40$ Identity matrix to make the Hamiltonian $256 \times 256$ and represent it on the quantum computer in terms of eight qubits. Running the VQE algorithm with the SLQP optimizer we find  the results in table 3  which is in close agreement with the value found from exact diagonalization.
    \begin{figure}
    \centering
    \begin{minipage}[b]{0.5\linewidth}
      \centering
      \includegraphics[width=\linewidth]{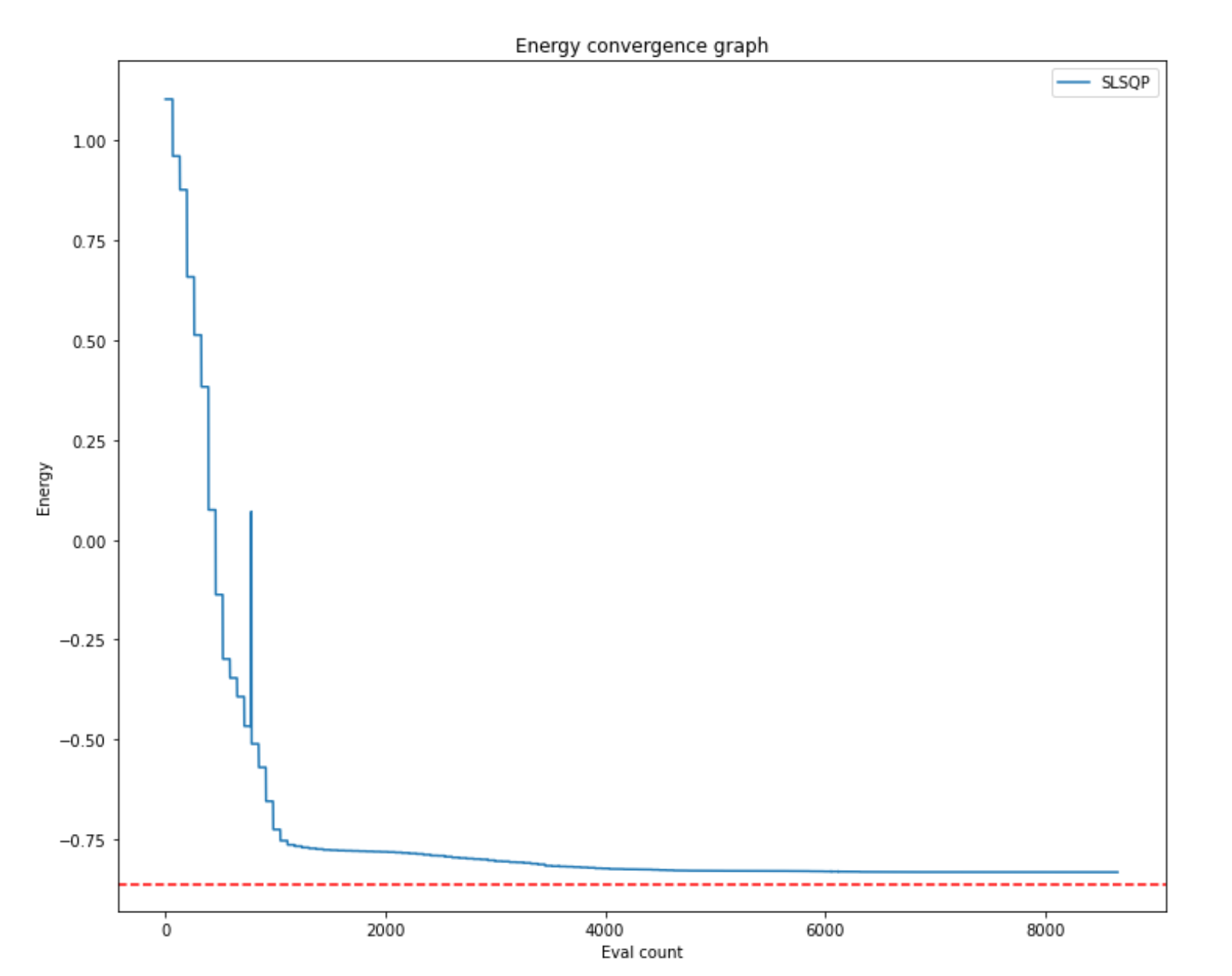}
    \end{minipage}
    \caption{Convergence graph for three boson links and three fermion vertices.}
    \label{3-qubit_unit}
\end{figure}
\begin{table}
\begin{tabular}{|l|l|l|l|l|}
\hline
Hamiltonian       & Qubits & Paulis  & Exact Result & VQE Result \\ \hline
Two $Z_3$ bosons Three fermions  & 8 & 547 &-0.86375772 & -0.83185930\\ 
 \hline
\end{tabular}
\caption{\label{tab:BasisCompare}  VQE results for $Z_3$ gauge model  model with three boson links and three fermion vertices with mass 0 and gauge coupling .15.  The Hamiltonian was mapped to 8-qubit operators with 547 Pauli terms.  The quantum circuit for each simulation utilized an \(R_y R_z\) variational form, with a fully entangled circuit of depth 3. The backend used was a statevector simulator. The Sequential Least SQuares Programming (SLSQP) optimizer was used, with a maximum of 600 iterations.}
\end{table}
\newpage
\section{Equation of state (EOS)  at finite density}

To determine the Equation of state for the $Z_3$ gauge model coupled to fermions we consider the three fermion vertex three gauge link case considered above and vary the chemical potential  at zero temperature. Then the estimate for the ground state energy obtained from the VQE at different chemical potentials will yield an equation of state as a function of chemical potential. The result of the eight qubit calculations are shown in figure 5. For this calculation we fixed the gauge coupling at $g=.15$ and mass $m=1.0$. The equation of state shows the correct behavior with energy rising linearly with the chemical potential. Note the VQE calculations are consistently above the exact values which is expected because the VQE algorithm determines an upper bound on the Energy for each value of the chemical potential. 
    \begin{figure}
    \centering
    \begin{minipage}[b]{1.0\linewidth}
      \centering
      \includegraphics[width=.8\linewidth]{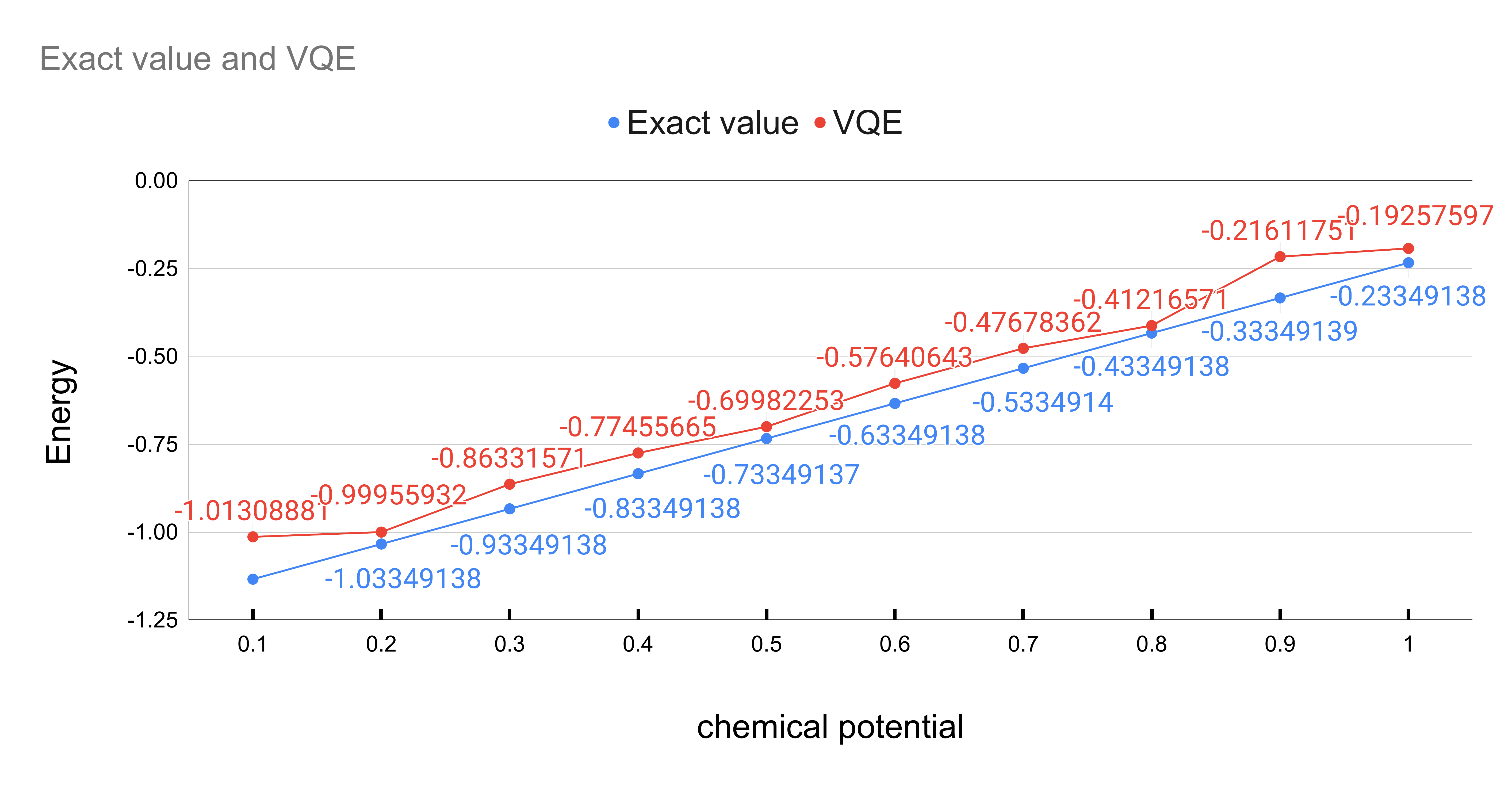}
    \end{minipage}
    \caption{Equation of State for $Z_3$ gauge theory coupled to fermions with nonzero chemical potential.}
    \label{3-qubit_unit}
\end{figure}

\newpage

\section{Conclusions}

In paper we studied  $Z_3$ gauge theory in 1+1 dimensions coupled to fermions. We used the hybrid classical-VQE quantum algorithm and obtain highly accurate results compared with exact diagonalization. We studied the theory at finite density to examine the equation of state for the model at finite chemical potential a quantity which is difficult to examine classically. 
Because of the presence of sign problems, finite density, non zero chemical potential  and time evolution the $Z_3$ gauge theory coupled to fermions is an excellent model to pursue quantum advantage over classical computers especially with  the increase in the number of qubits and improved development of noise mitigation techniques. It will also be interesting to explore the finite temperature equation of state for this model as well perhaps using a thermo-double approach to realizing finite temperature on a quantum computer
\cite{Miceli:2019sym}
\cite{Cottrell:2018ash}
\cite{Wu:2018nrn}
\cite{Swingle:2016foj}. Finally it will be interesting see if the variational Schrodinger wave function approach to higher dimensional QCD \cite{Ebadati:2018pmq}
\cite{Feuchter:2004mk}
\cite{Feynman:1981ss}
\cite{Cornwall:2007dh}
\cite{Samuel:1996bt}
\cite{Sala:2018dui} can realized in terms of quantum computing by adapting  the classical variational techniques to the quantum computing variational approaches in terms of gates on near term quantum hardware.

\section*{Acknowledgements}

 This material is based upon work supported in part by the U.S. Department of Energy, Office of Science, National Quantum Information Science Research Centers, Co-design Center for Quantum Advantage (C2QA) under contract number DE-SC0012704. This project was supported in part by the U.S. Department of Energy, Office of Science, Office of Workforce Development for Teachers and Scientists (WDTS) under the Science Undergraduate Laboratory Internships Program (SULI). We wish to acknowledge useful discussions on $Z_3$ gauge models with Rob Pisarski and Yannick Meurice and wish to thank Rob Pisarski for suggesting this problem to us.

\end{document}